\documentclass[sigconf,pbalance]{acmart}
\AtBeginDocument{%
  }

\setcopyright{acmlicensed}

\copyrightyear{2025}
\acmYear{2025}
\setcopyright{acmlicensed}
\acmConference[MM '25] {Proceedings of the 33rd ACM International Conference on Multimedia}{October 27--31, 2025}{Dublin, Ireland.}
\acmBooktitle{Proceedings of the 33rd ACM International Conference on Multimedia (MM '25), October 27--31, 2025, Dublin, Ireland}
\acmISBN{979-8-4007-2035-2/2025/10}
\acmDOI{10.1145/3746027.3754576}
\usepackage[ruled,linesnumbered]{algorithm2e}
\usepackage{colortbl, xcolor} 

\usepackage{multicol}
\usepackage{dsfont}
\usepackage{multirow}
\usepackage{amsmath}
\usepackage{enumitem}
\usepackage{eucal}
\usepackage{bm}
\usepackage{amsfonts}
\usepackage{subfigure}
\usepackage[switch]{lineno}
\usepackage{bbding}
\usepackage{listings}
\usepackage{balance}
\usepackage{makecell}

\begin{document}

\title{Dark Side of Modalities: Reinforced Multimodal Distillation for Multimodal Knowledge Graph Reasoning}

\author{Yu Zhao}
\email{zhaoyu@dbis.nankai.edu.cn}
\affiliation{%
  \institution{VCIP, DISSec, College of Computer Science\\Nankai University}
  \city{Tianjin}
  \country{China}
}

\author{Ying Zhang}
\authornote{Corresponding author.}
\email{yingzhang@nankai.edu.cn}
\affiliation{%
  \institution{VCIP, DISSec, College of Computer Science\\Nankai University}
  \city{Tianjin}
  \country{China}
}

\author{Xuhui Sui}
\email{suixuhui@dbis.nankai.edu.cn}
\affiliation{%
  \institution{VCIP, DISSec, College of Computer Science\\Nankai University}
  \city{Tianjin}
  \country{China}
}

\author{Baohang Zhou}
\email{zhoubaohang@dbis.nankai.edu.cn}
\affiliation{%
  \institution{School of Software\\Tiangong University}
  \city{Tianjin}
  \country{China}
}

\author{Haoze Zhu}
\email{zhuhaoze@dbis.nankai.edu.cn}
\affiliation{%
  \institution{VCIP, DISSec, College of Computer Science\\Nankai University}
  \city{Tianjin}
  \country{China}
}

\author{Jeff Z. Pan}
\email{j.z.pan@ed.ac.uk}
\affiliation{%
  \institution{School of Informatics\\University of Edinburgh}
  \country{Edinburgh, UK}
}

\author{Xiaojie Yuan}
\email{yuanxj@nankai.edu.cn}
\affiliation{%
  \institution{VCIP, DISSec, College of Computer Science\\Nankai University}
  \city{Tianjin}
  \country{China}
}


\begin{abstract}
The multimodal knowledge graph reasoning (MKGR) task aims to predict the missing facts in the incomplete MKGs by leveraging auxiliary images and descriptions of entities. 
Existing approaches are trained with single-target objectives, which neglect the probabilistic correlations of entity labels, especially in non-target entities. 
Moreover, previous studies incorporate all modalities statically or adaptively, overlooking the negative impacts of irrelevant or misleading information in the incompetent modalities.
To address these issues, we introduce a novel Reinforced Multimodal Distillation framework, exploiting the \textit{Dark Side of Modalities} (\textbf{DSoM}) from two perspectives: 
\textbf{(1) Dark knowledge from non-target entities}: We propose to train a \textit{unimodal KGR model} through logit distillation to mimic the multimodal soft labels provided by pre-trained multimodal teacher models. 
The multimodal soft labels could provide rich supervision signals with subtle correlations among both target and non-target entities from multiple perspectives. 
We further decouple logits into neighbor entities and non-neighbor entities to divide into two types of correlations. 
\textbf{(2) Dark side in unhelpful modalities}: To exclude the adverse effects of unhelpful modalities, we introduce a reinforced teacher combination mechanism that dynamically selects the optimal set of multimodal teachers for each triple. 
The agent is trained to maximize the rewards, which are only assigned to the beneficial multimodal combination strategies for the student model. 
Comprehensive experiments demonstrate the effectiveness of DSoM framework on 5 MKGR datasets. 
Codes are available at \href{https://github.com/OreOZhao/DSoM}{github.com/OreOZhao/DSoM}.
\end{abstract}






\renewcommand{\shortauthors}{Yu Zhao et al.}

\maketitle

\section{Introduction}
\begin{figure}[!t]
    \centering
    \includegraphics[width=0.48\textwidth]{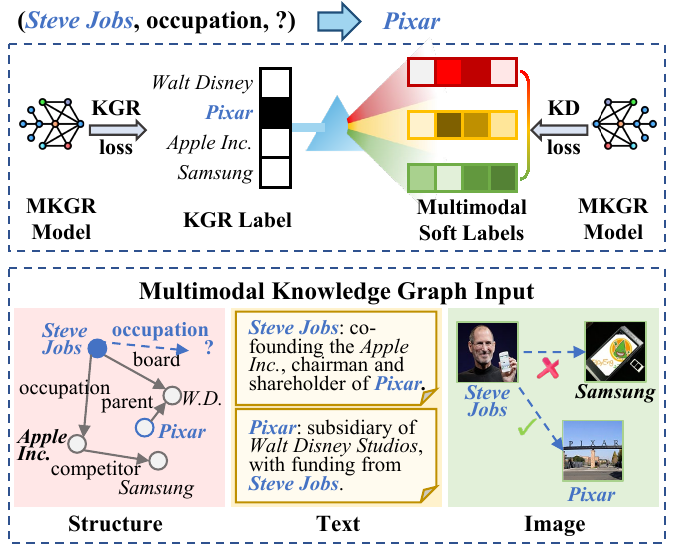}
    \vspace{-15pt}
    \caption{We propose to train the MKGR model with (1) multimodal soft labels to exploit dark knowledge, (2) reinforced multimodal selection to exclude unhelpful modalities.}
    \label{fig:intro}
    \vspace{-25pt}
    
\end{figure}

Multimodal Knowledge Graphs (MKGs) \cite{zhu2022mmkgsurvey, Liangke2022mmkgSurvey, chen2024KGmeetMMSurvey} organize the relational triples \textit{(head entity, relation, tail entity)} in KGs with their multimodal information, including images and descriptions of entities. 
MKGs have been applied to knowledge-intensive tasks such as Recommender Systems \cite{sun2020mmkgrs}, Visual Question Answer \cite{marino2019okvqa}, Information Extraction \cite{www25zhou}, etc. 
Since MKGs are far from complete, due to possible unobserved facts or knowledge emergence, the Multimodal Knowledge Graph Reasoning (MKGR) \cite{xie2016DKRL,xie2017IKRL} task is proposed to exploit the multimodal information to predict the missing facts in MKGs. 
For example, given the query \textit{(Steve Jobs, occupation, ?)} in Figure \ref{fig:intro}, the MKGR task aims to predict the tail entity \textit{Pixar} based on structural, textual, and visual modalities.

The MKGR task has been widely explored \cite{mousselly2018TBKGC,wang2021rsme,xu2022MMRNS,zhang2024mygo,liang2023SGMPT,FedMKGC2025zhaoyu,www25sui} in multiple perspectives. 
They propose static fusion \cite{xie2017IKRL,mousselly2018TBKGC,wang2019transAE,lu2022mmkrl,zhang2023MANS,zhang2022VBKGC,zhang2023maco} or adaptive fusion \cite{wang2021rsme,cao2022otkge,xu2022MMRNS,zhang2024unleashing,lee2023vista,zhang2024mygo,li2023imf} mechanisms to incorporate multimodal features or dynamically ensemble the multimodal predictions \cite{zhao2022mose,li2023imf}.
Moreover, they usually optimize their multimodal models with the cross-entropy objectives \cite{zhao2022mose,cao2022otkge,li2023imf,zhang2024mygo} or margin-ranking objectives \cite{xie2017IKRL,mousselly2018TBKGC,xu2022MMRNS,zhang2024unleashing}. 
The cross-entropy objective optimized methods formulate the MKGR task as an entity classification task, where only the probability of the target entity is maximized.
The margin-ranking objective optimized methods aim to rank the score of the target entity higher than those of negative entities and maximize their distance.
Both objectives treat the negative entities uniformly. 

Despite their great progress, they overlooked the Dark Side of Modalities in MKGR in two aspects: 
\textbf{(1) \textit{Dark knowledge} in non-target entities, where multi-modalities could provide multiple perspectives. }
First, the inter-entity label correlations are systematically neglected in existing cross-entropy or margin-ranking optimized MKGR methods. 
As shown in Figure \ref{fig:intro}, for triple \textit{(Steve Jobs, occupation, Pixar)}, existing approaches only emphasize the target entity \textit{Pixar} in the one-hot hard labels, while both other true answers such as \textit{Apple Inc.} are neglected. 
In addition, the probability correlations of entities, such as ranking \textit{Pixar} over \textit{Walt Disney} and \textit{Apple Inc.} over \textit{Samsung}, could also provide supervision to improve distinguishing and generalization ability.
Inspired by Knowledge Distillation (KD) research \cite{hinton2015KD}, the soft labels, i.e., continuous probability distributions over all entity labels, could provide rich so-called \textbf{\textit{dark knowledge}} as supervision signals. 
Moreover, \textbf{multimodal soft labels could capture the entity probability correlations from different perspectives.}
Since modalities usually have diverse knowledge coverage, they exhibit varying relevance levels between the query and all entities. 
For example, based on the structural linkages and entity descriptions, the structural and textual modalities gave both true entities \textit{Pixar} and \textit{Apple Inc.} high probabilities.
The multimodal soft labels can thus provide more comprehensive and generalizable supervision signals than the traditional one-hot labels \cite{2020understanding-KD-ensemble}, 
analogous to how white light is dispersed into different colors of light through a prism in Figure \ref{fig:intro} where each color of light represents one modal of dark knowledge.

\textbf{(2) \textit{Dark side} in unhelpful modalities that have negative impacts on reasoning. }
Existing approaches \cite{wang2021rsme,zhao2022mose,chen2022mkgformer} have investigated the multimodal contradictions in MKGR, i.e., the relevance of one modality may contradict the other since different modalities consist of different perspectives of knowledge.
For example in Figure \ref{fig:intro}, the graph structures and entity descriptions can point to true entity \textit{Pixar}, while the images show more relevant information between \textit{Steve Jobs} and non-target entity \textit{Samsung} for having the similar Phone component in pictures. 
Existing approaches proposed to lower the noisy modal weights through fusion gates \cite{wang2021rsme}, attention mechanisms \cite{xie2017IKRL,lee2023vista,zhang2024mygo}, and decision ensemble methods \cite{zhao2022mose}. 
These methods inherently assume that all modalities should be incorporated, including the negative ones. 
However, even given a low weight, the unhelpful modalities would provide misleading signals for model optimization, which is the \textbf{\textit{dark side}} of MKGR. 
Here, we hypothesize that the \textbf{{discrete decision} to entirely exclude the negative modalities for corresponding triples would thoroughly prevent their negative impacts}. 
Moreover, the multimodal soft labels can reflect the knowledge certainty and quality of modalities for supporting the decision.

Based on the above insights, we propose a Reinforced Multimodal Distillation framework for Multimodal Knowledge Graph Reasoning to exploit the Dark Side of Modalities (\textbf{DSoM}).
Specifically, we propose to train a unimodal student model to mimic the multimodal soft labels from multimodal teachers, which are dynamically combined through a reinforcement procedure. 
\textbf{(1) Reinforced multimodal teacher combination}: 
We design an agent to dynamically choose the optimal set of multimodal teachers for the student and exclude the unhelpful ones. 
We first pre-train the multimodal teachers to obtain the multimodal soft-labels. 
The agent then samples the multimodal combination strategy based on the state composed of multimodal soft labels.
The agent rewards the beneficial strategies that obtain better performance than the student model, and penalizes the opposite ones. 
The policy network is optimized toward maximizing rewards. 
\textbf{(2) Neighbor-decoupled knowledge distillation:} 
We propose a neighbor-decoupled knowledge distillation objective to guide the student model with soft labels of combined multimodal teachers.
Since in MKGs, the query usually consists of multiple true targets, i.e., neighboring entities such as \textit{Apple Inc.} and \textit{Pixar} in Figure \ref{fig:intro}, 
we propose to decouple the distillation into neighbor and non-neighbor entity KD inspired by \cite{zhao2022dkd}. 
This way, the student could separately mimic the logits from two types of correlations among true entities and wrong entities. 
Comprehensive experiments show that DSoM achieves state-of-the-art MKGR performance on 5 datasets, with detailed analyses further highlighting the effectiveness and efficiency of DSoM. 
The contributions of our paper can be summarized as follows:
\begin{itemize}
    \item As far as we know, we are the first MKGR research to train a unimodal model with multimodal soft labels via knowledge distillation, 
    rather than a multimodal model with one-hot hard labels, 
    aiming to exploit dark knowledge of rich entity label correlations from multimodal perspectives.  
    \item We are the first MKGR research to discretely select sample-wise optimal modality set via the reinforced combination mechanism, rather than incorporating all modalities, aiming to exclude the sample-wise unhelpful modalities. 
    \item Experiments demonstrate the effectiveness of our proposed DSoM. Detailed studies validate the necessity and efficiency of exploiting the dark side of modalities in two folds.
\end{itemize}

\section{Related Work} \label{sec: related work}

\subsection{Multimodal Knowledge Graph Reasoning}
Multimodal knowledge graph reasoning (MKGR) \cite{zhu2022mmkgsurvey, Liangke2022mmkgSurvey, chen2024KGmeetMMSurvey} aims to exploit multimodal information, including entity images \cite{xie2017IKRL} and entity descriptions \cite{xie2016DKRL}, to enhance the knowledge graph reasoning performance in predicting missing facts. 
Existing multimodal approaches \cite{xie2017IKRL,wang2021rsme,cao2022otkge,xu2022MMRNS,zhang2024unleashing,zhang2024mygo,liu2024dysarl} typically extend the unimodal knowledge graph embedding (KGE) \cite{bordes2013TransE,trouillon2016complex,sun2018RotatE,balazevic2019tucker} approaches to multimodal settings. 
The unimodal KGE approaches, such as TransE \cite{bordes2013TransE}, ComplEx \cite{trouillon2016complex}, RotatE \cite{sun2018RotatE}, TuckER \cite{balazevic2019tucker}, define the score functions that measure the plausibility score of the triples. 
The defined score functions are exploited in (1) the training stage, to optimize the entity and relation embeddings to give the observed triples higher scores, and (2) the testing stage, to rank the candidate entities throughout the KG for evaluation. 
These KGE approaches only model the structural modality, neglecting the rich and helpful semantics in visual and textual modalities. 

Existing MKGR studies extend the unimodal KGE approaches in the following aspects:
\textit{(1) Multimodal alignment}: 
they \cite{xie2017IKRL,mousselly2018TBKGC,wang2023QEB,lu2022mmkrl,zhang2022VBKGC,ACL23TEA,IPM2025ME3A} project the multimodal embeddings in the structural representation space and design score functions to align them. 
OTKGE \cite{cao2022otkge} proposes multimodal distribution-view alignment with optimal transport. 
\textit{(2) Multimodal fusion}: they exploit mechanisms including auto-encoders \cite{wang2019transAE}, gate functions \cite{wang2021rsme}, bilinear models \cite{li2023imf}, and transformers \cite{zhang2022VBKGC,lee2023vista,zhang2024mygo} to fuse the multimodal embeddings. 
\textit{(3) Negative sampling}: they \cite{xu2022MMRNS,zhang2023MANS,zhang2024unleashing} focus on enhancing the quality of negative samples with multimodal information. 
\textit{(4) Multi-hop reasoning}: They \cite{liu2024dysarl,zheng2025TKDE_MKGR,zheng2023icde_mkgr} exploit deep structure-based methods to model the multimodal graph structures in MKGs. 
\textit{(5) Decision ensemble}: they exploit boosting algorithm \cite{zhao2022mose}, meta weight learning \cite{zhao2022mose}, weighted parameters \cite{li2023imf}, or balance factors \cite{xu2022MMRNS} to fuse the multimodal prediction results. 
There are also pre-trained language/multimodal model enhanced approaches with transformer architectures \cite{chen2022mkgformer,liang2023SGMPT,zhao2024CMR}, which also show promising performance. 

Our paper can be categorized as a decision ensemble approach for assembling multimodal teacher predictions in a student model. 
Moreover, the distilled student model learns from beneficial essence in multimodal teachers and is inherently composed of multimodal semantics, thus our paper could also be seen as a multimodal fusion approach.
However, the key insights of our paper lie in designing a framework to learn from (1) dark knowledge in multimodal soft labels that reveal the entity label correlations, and (2) selective multimodal combination for eliminating negative impacts of unhelpful modalities, which are not explored by existing studies.

\subsection{Knowledge Distillation for KGs}
Knowledge Distillation (KD) \cite{hinton2015KD} was proposed earlier for model compression, which consists of a larger teacher model and a smaller student model.
By forcing the student model to mimic the soft logits from the teacher model, the student model can learn from ``dark knowledge'' from soft labels and rapidly converge to similar performance with the teacher model. 
KD has been widely studied in recent years and can be categorized into (1) logit distillation \cite{zhang2018mutual-logit-distillation,mirzadeh2020TA-KD,zhao2022dkd,jin2023multi-level-logit-distil}, which minimizes the KL divergence between logits of teacher and student, 
and (2) feature distillation \cite{romero2014Fitnets-feature-distill,heo2019comprehensive-Feature-distill,park2019relational-feature-KD}, which minimizes the feature distance between teacher and student. 
Since we mainly focus on designing objectives with multimodal soft labels for the MKGR model, we mainly focus on the logit distillation. 

KD is applied in KGR for compressing unimodal KGE models \cite{wang2021MulDE,zhu2022DualDE,liu2023IterDE,yu2023WTT-KD,fan2024PMD-PLM-KGC-KD,liu2024CSD,xu2024ka-dtd} or network pruning \cite{li2023MetaSD-prune} for efficiency. 
These approaches mainly aim to distill knowledge from a higher-dimensional embedding KGE model \cite{wang2021MulDE,zhu2022DualDE,liu2023IterDE,yu2023WTT-KD}, a more-layer transformer encoder model \cite{fan2024PMD-PLM-KGC-KD}, or a unpruned source model \cite{li2023MetaSD-prune}, to a smaller student model. 
MulDE \cite{wang2021MulDE} proposes a multi-teacher KD framework, where the teachers are different KGE teachers with different score functions, which is the only existing multi-teacher KD-KGR framework. 
Other approaches improve the KD-KGR with dual-stage distillation \cite{zhu2022DualDE}, iterative distillation \cite{liu2023IterDE}, dynamic soft label weights \cite{zhu2022DualDE,liu2023IterDE,yu2023WTT-KD}, dynamic temperature \cite{liu2024CSD,xu2024ka-dtd}. 
They mainly focus on unimodal-to-unimodal distillation on unimodal KGR, while our paper focuses on multimodal-to-unimodal distillation for multimodal KGR to exploit the dark knowledge in modalities.
We propose reinforced multimodal teacher combination and neighbor-decoupled KD mechanisms for MKGR.

\begin{figure*}[!t]
    \centering
    \includegraphics[width=\textwidth]{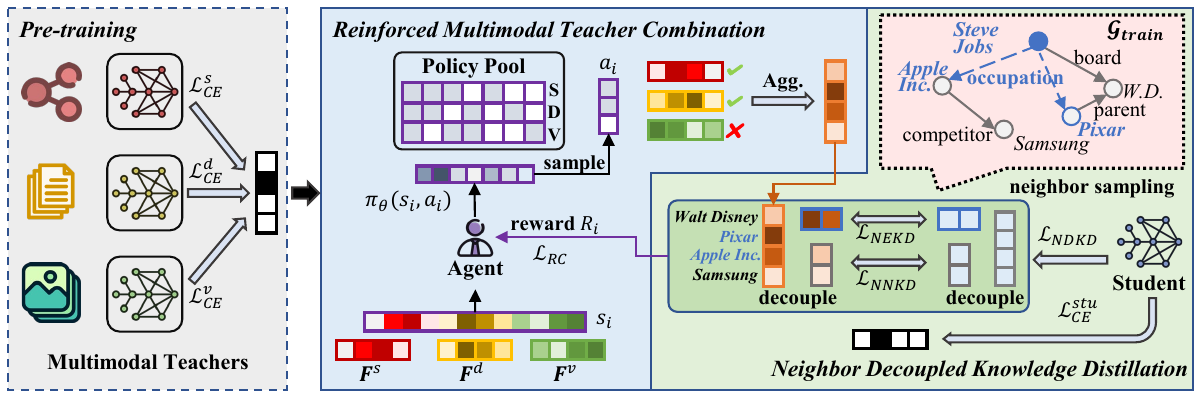}
    \vspace{-12pt}
    \caption{The DSoM framework for MKGR. First, we pre-train the multimodal teachers. Then we propose the reinforced multimodal teacher combination module to select sample-wise helpful teachers. 
    We propose the neighbor-decoupled knowledge distillation to guide the student in learning from neighbor and non-neighbor correlations in mimicking teacher logits. }
    \label{fig:method}
    \vspace{-12pt}

\end{figure*}

\section{Preliminaries}
\subsection{Multimodal Knowledge Graph Reasoning}

\textbf{Multimodal knowledge graphs} (MKGs) are knowledge graphs with multimodal information, including structural triples, entity images and entity descriptions.
An MKG could be denoted as $\mathcal{G} = \{\mathcal{E}, \mathcal{R}, \mathcal{T}, \mathcal{V}, \mathcal{D} \}$, where $\mathcal{E}$ is the set of entities, $\mathcal{R}$ is the set of relations, $\mathcal{T}$ is the set of triples $\{(h,r,t)\}\subseteq \mathcal{E}\times\mathcal{R}\times\mathcal{E}$, and $\mathcal{V}, \mathcal{D}$ is the set of entity images and descriptions. 
We denote the multimodal embeddings of entities and relations following modality-split studies \cite{zhao2022mose,li2023imf} as $\mathbf{E}^m=[e^m] \in \mathbb{R}^{|\mathcal{E}|\times dim}$ and $\mathbf{R}^m=[r^m] \in \mathbb{R}^{|\mathcal{R}|\times dim}$. The $m\in\mathcal{M}=\{\mathcal{S,V,D}\}$ denotes structural, visual, and textual modalities and $dim$ denotes embedding dimension.
The visual and textual entity embeddings are initialized with projected multimodal features $E^v$ and $E^d$ as $\mathbf{E}^v=\mathbf{W}^v E^v$ and $\mathbf{E}^d=\mathbf{W}^d E^d$ from pretrained visual and language models \cite{dosovitskiy2020ViT,devlin2019bert}, where $E^v,E^d$ are fixed and the projection weights $\mathbf{W}^v,\mathbf{W}^d$ are trainable. 
The structural entity embeddings and all the relation embeddings are randomly initialized.

\textbf{Multimodal Knowledge Graph Reasoning} (MKGR) aims to predict the missing entity with multimodal information. 
For a particular query $q = (h,r,?)$ in the training triples $\mathcal{T}_{train}$, MKGR model aims to ranking all the candidate entities $e\in\mathcal{E}$ based on the score distribution $\mathbf{F}(h,r,t)=[f(h,r,e;\mathbf{E},\mathbf{R})]\in \mathbb{R}^{|\mathcal{E}|}, e \in \mathcal{E}$ and make the target entity $t$ the highest among other entities. 
MKGR models are typically optimized to minimize the cross-entropy loss \cite{zhao2022mose,li2023imf,zhang2024mygo} as Equation \eqref{eq:ce KGR}, where $\sigma$ denotes softmax function. 
The reversed triple $(t, r^{-1}, h)$ are also added for unifying the head prediction and tail prediction into tail prediction \cite{bordes2013TransE}.
The score functions for MKGR \cite{xie2017IKRL,zhao2022mose,zhang2024unleashing,li2023imf} can be derived from KGE functions \cite{bordes2013TransE,trouillon2016complex,sun2018RotatE,balazevic2019tucker}.
\begin{equation} \label{eq:ce KGR}
\small
\begin{aligned}
    \mathcal{L}_{CE}&= -\frac{1}{|\mathcal{T}_{train}|} \sum_{(h,r,t)\in \mathcal{T}_{train}} \log(\sigma(\mathbf{F}(h,r,t)))\\
    &=CE (\sigma(\mathbf{F}(h,r,t)))
\end{aligned}
\end{equation}

\subsection{Vanilla Knowledge Distillation}
\textbf{Knowledge distillation} (KD) \cite{hinton2015KD} is proposed as a model compression framework to transfer the knowledge in the larger teacher model to the smaller student model. 
Specifically, 
the KD process optimizes the student model to mimic the teacher's logit outputs by minimizing the KL divergence between two logits.
Take a $C$-way classification task as an example, the vanilla KD loss is shown in Equation \eqref{eq: KD}, where $\mathbf{P} = [{p}_i] = [\sigma(z_i/\tau)]\in\mathbb{R}^{|C|}$ is the temperature $\tau$-scaled class probability of class logit $z_i$ of a C-class probability of a single input. The temperature scaling with $\tau$ softens the original logits to avoid the overconfidence phenomenon. 
\begin{equation} \label{eq: KD}
\small
    \mathcal{L}_{KD}= {KL}(\mathbf{P}^{tea}\parallel\mathbf{P}^{stu}) = \sum_{i=1}^C \mathbf{P}^{tea}_i \log (\frac{\mathbf{P}^{tea}_i}{\mathbf{P}^{stu}_i})
\end{equation}

\section{Methodology}

\subsection{Overview}
Figure \ref{fig:method} shows the framework of DSoM. 
First, we propose to pre-train the multimodal teachers simultaneously with respective cross-entropy loss. 
The student model is trained with both cross-entropy loss with hard labels and the logit distillation loss with multimodal soft labels. 
We propose the reinforced multimodal teacher combination module to eliminate the modalities with negative impacts. 
Specifically, the agent selects an optimal policy from the policy pool based on the state composed of soft labels of multimodal teachers. 
The policy network is trained using the rewards derived from the performance of the combined teachers. 
Moreover, we propose the neighbor-decoupled knowledge distillation objective to guide the student model in learning from dark knowledge of two types of entity correlations. 
The logits of combined teachers and the student are decoupled into neighbor entities and non-neighbor entities. 
Then we train the student MKGR model to mimic the logits of two terms separately. 
At test time, the student model is used for evaluation.

\subsection{Multimodal Teacher Model Pre-training}
We pre-train the multimodal teachers in a multimodal-split manner \cite{zhao2022mose,li2023imf}, where the entity $\mathbf{E}^m$ and relation $\mathbf{R}^m$ representations for each modality, as well as the training objectives, are split. 
The pre-training objective is $\mathcal{L}^{tea}$ as Equation \eqref{eq:teacher pretraining}. 
This way, each modality can output its logits based on its knowledge coverage, quality, relevance, and so on. 
In our paper, we follow \cite{zhao2022mose} to conduct experiments with ComplEx \cite{trouillon2016complex} as the score function. 
\begin{equation} \label{eq:teacher pretraining}
\small
\begin{aligned}
        \mathcal{L}^{tea} = \sum_{m\in\mathcal{M}} \mathcal{L}^m_{CE} =&\sum_{m\in\mathcal{M}}CE (\sigma(\mathbf{F}^m(h,r,t)))\\
        =&\sum_{m\in\mathcal{M}}CE (\sigma([f(h,r,e;\mathbf{E}^m,\mathbf{R}^m)]))
\end{aligned}
\end{equation}

\subsection{Unimodal Student Model Training}
\textbf{Student KGR Objective.} The student MKGR model is randomly initialized in the same way as the unimodal KGR model, and mainly exploits the multimodal teacher supervision signals to integrate the multimodal information. 
The student model itself does not involve multimodal information or features. 
The KGR objective $\mathcal{L}_{CE}^{stu}$ of the student model is as Equation \eqref{eq:ce KGR}, where $F^{stu}(h,r,t)=[f(h,r,e;\mathbf{E}^{stu},\mathbf{R}^{stu})]$, and 
$\mathbf{E}^{stu},\mathbf{R}^{stu}$ is the entity and relation embeddings of the student model. 
At test time, only the entity and relation embeddings of the student model are used for evaluation.

\textbf{Student Overall Objective.} The overall objective for the student training stage is as Equation \eqref{eq:overall stu}, where $\mathcal{L}_{RC}$ and $\mathcal{L}_{NDKD}$ represent the loss of reinforced multimodal teacher combination and that of neighbor-decoupled knowledge distillation, respectively. 
We exploit a hyper-parameter $\gamma$ to adjust the weight of $\mathcal{L}_{NDKD}$. 
\begin{equation} \label{eq:overall stu}
\small
    \mathcal{L}^{stu} = \mathcal{L}_{CE}^{stu} + \mathcal{L}_{RC} + \gamma\mathcal{L}_{NDKD}
\end{equation}

\subsection{Reinforced Teacher Combination}
Though multimodal teachers are pre-trained with the training set, they still face multimodal contradictions in that not all modalities are effective for all triples due to modality quality, relevance to the triple, and modality knowledge coverage. 
Thus, we propose the reinforced multimodal teacher combination module to rule out the negative impacts of the unbeneficial modalities. 
In each iteration, an agent interacts with the environment consisting of multimodal teachers and receives the logits of multimodal teachers as the current state. 
Then, the agent samples the multimodal teachers based on the combination policy probability and aggregates the combined teachers to participate in the following distillation stage. 
The reward is calculated based on the performance of aggregated teachers and used to update the policy network.

\textbf{State. }
We design the state $s_i=s_{(h,r,t)}$ of the $i$-th triple $(h,r,t)$ as the concatenation of the score distributions $\mathbf{F}^m(h,r,t)$ from Equation \eqref{eq:teacher pretraining} of all teachers.
In our paper, DSoM models structural, visual, and textual modal scores and can be extended to any modalities. 
\begin{equation}\label{eq:state}
\small
s_i=s_{(h,r,t)} = [\mathbf{F}^s(h,r,t)\| \mathbf{F}^v(h,r,t)\|\mathbf{F}^d(h,r,t)] 
\end{equation}

\textbf{Action. }
We design an agent to select a policy from the policy pool that consists of all possible combinations of all modalities. Notably, for excluding the situation where none of the teachers are selected, the policy pool of DSoM is sized as $2^{|\mathcal{M}|}-1$. 
As shown in Figure \ref{fig:method}, the policy pool includes $7$ situations for our modality setting, including structural, textual, and visual modalities. The painted and unpainted blocks represent selected and unselected for the modal teacher. 
The agent chooses from the policy pool $\mathcal{P}$ that determines which of the teachers are selected. 
As Equation \eqref{eq: action}, a policy function $\pi_\theta(s_i,a_i)$ outputs the distribution conditioned on the states, from which the action $a_i \in \mathcal{P}$ is sampled. 
We implement the policy function $\theta$ with a trainable Multilayer Perceptron activated by ReLU, denoted as $\mathbf{W}_p$. 
The parameter of policy function $\theta$ projects the state $s_i$ sized $|\mathcal{M}|\cdot|\mathcal{E}|$ to the policy pool sized $2^{|\mathcal{M}|}-1$.
\begin{equation} \label{eq: action}
\small
   a_i \sim \pi_\theta(s_i,a_i) = \sigma(\mathbf{W}_p(s_i)) 
\end{equation}

We denote the chosen teachers of sampled action $a_i$ as $\mathcal{A}_i\subseteq\mathcal{M}$. 
The logits of teachers are then aggregated by a simple mean procedure as Equation \eqref{eq: tea logits}.
The trainable parameters of reinforced combined teacher logits are $\theta$ in the policy function. 
\begin{equation} \label{eq: tea logits}
\small
\mathbf{F}^{tea}(h,r,t)= \frac{1}{|\mathcal{A}_i|}\sum_{m}^{\mathcal{A}_i}\mathbf{F}^{m}(h,r,t) = [f^{tea}(h,r,e;\theta)]
\end{equation}

\textbf{Reward. }
The goal of the agent is to select the best multimodal teacher combination policy for each triple to better guide the student. 
Thus, we design a simple reward function to reward the action $a_i$ that selects beneficial teachers for the student, which means the cross-entropy of selected teachers $\mathcal{A}_i$ is lower than that of the student.
As Equation \eqref{eq:reward}, the reward $R_i$ is a positive value if the selected teachers are better than the student, and a negative value vice versa. 
\begin{equation} \label{eq:reward}
\small
  R_i = \begin{cases}
1, \textit{\quad if\quad}CE(\sigma(\mathbf{F}^{tea}(h,r,t))) <CE(\sigma(\mathbf{F}^{stu}(h,r,t))) \\
-10, \textit{elsewise. }
\end{cases}
\end{equation}

\textbf{Optimization. }
The optimization of the agent aims to find the parameter $\theta$ with a larger reward. 
We use the policy gradient algorithm following \cite{williams1992reinforcegradient} as Equation \eqref{eq:gradientpolicy}, where $\nabla \theta$ represents the gradient at $i$-step of an episode. 
The $\alpha$ is the learning step size. 
Moreover, we exploit a baseline function for reducing the variance, where $\delta=(R_i - \hat{R}_i)$ and $\hat{R}_i$ is the baseline reward gained by averaging logits of all teachers. 
\begin{equation}\label{eq:gradientpolicy}
    \nabla \theta = \alpha \delta \nabla \log \pi_\theta(s_i,a_i)
\end{equation}
The loss for the reinforced multimodal teacher combination can be formulated as Equation \eqref{eq:RC} for the policy function to assign a larger probability to the policy with more rewards.
\begin{equation} \label{eq:RC}
\small
\mathcal{L}_{RC} = -\frac{1}{|\mathcal{T}_{train}|} \sum_{(h,r,t)\in \mathcal{T}_{train}}\delta\log \pi_\theta(s_i,a_i)
\end{equation}

\subsection{Neighbor Decoupled Knowledge Distillation}
After obtaining the teacher logits $\mathbf{F}^{tea}(h,r,t)$ from our proposed reinforced combination, the student could learn to mimic the teachers with the knowledge distillation objective. 
For distinguishing the correlations of neighbors, including all true answer entities of the query, and those of non-neighbors, inspired by \cite{zhao2022dkd}, we propose a neighbor-decoupled KD objective. 

First, we obtain the temperature-scaled probability distribution of teacher logits and student logits as Equation \eqref{eq:tea stu prob}.
Each probability $p_e$ represents the probability that entity $e\in\mathcal{E}$ is the target entity of query $(h,r,?)$ in the combined teacher logits and student logits, respectively.
\begin{equation}\label{eq:tea stu prob}
\small
\begin{aligned}
\mathbf{P}^{tea}(h,r,t)&  = \sigma(\mathbf{F}^{tea}(h,r,t)/\tau)  = [p_e^{tea}]\\
\mathbf{P}^{stu}(h,r,t)& = \sigma(\mathbf{F}^{stu}(h,r,t)/\tau) = [p_e^{stu}]
\end{aligned}
\end{equation}

The original Decoupled KD \cite{zhao2022dkd} decouples the logits of the target class and non-target classes as $\mathcal{L}_{DKD}$ in Equation \eqref{eq:dkd}.
Different with vanilla KD, the all entity distributions are decoupled to (1) the binary probability $\mathbf{b} =[p_t, 1-p_t]$ of target entity $t$, (2) non-target entities probabilities $\hat{p}_e =p_e/p_{\backslash t}$ with $e\in\mathcal{E}-t$, for further exploiting the dark knowledge in non-target labels. 
However, in the MKGR task, there are usually multiple neighboring entities for a query $(h,r,?)$, which are all true answers to the query.
We define the neighbor set of the query as $N(h,r)=\{n|(h,r,n)\in\mathcal{T}_{train}\}$, 
which includes the target entity $t$ of the specific triple and other true answers to the query. 
Thus, we extend the DKD into the neighbor-decoupled knowledge distillation and propose $\mathcal{L}_{NDKD}$ as Equation \eqref{eq:dkd}. 
The $\overline{\mathbf{b}} =[\overline{p}_n, 1-\overline{p}_n]$ of $\mathcal{L}_{NDKD}$ represents the average binary probability of all neighbor entities $\overline{p}_n = \frac{1}{|N|}\sum_n^N p_n$. 
And $\tilde{p}_e =p_e/p_{\backslash N}$ represents the probabilities of non-neighbor entities $e\in\mathcal{E}-N$.
\begin{equation}\label{eq:dkd}
\small
\begin{aligned}
&\mathcal{L}_{DKD} &&= \alpha KL(\mathbf{b}^{tea}\parallel\mathbf{b}^{stu}) + \beta KL(\hat{\mathbf{P}}^{tea}\parallel\hat{\mathbf{P}}^{stu})\\
&\mathcal{L}_{NDKD} &&=  \alpha KL(\overline{\mathbf{b}}^{tea}\parallel\overline{\mathbf{b}}^{stu}) + \beta KL(\tilde{\mathbf{P}}^{tea}\parallel\tilde{\mathbf{P}}^{stu})\\
& &&=\alpha \mathcal{L}_{NEKD} +\beta \mathcal{L}_{NNKD}
\end{aligned}
\end{equation}
This way, we decouple the knowledge distillation between reinforced combined teachers and the student into the KD of neighbor entities $\mathcal{L}_{NEKD}$ and KD of non-neighbor entities $\mathcal{L}_{NNKD}$. 
We follow \cite{zhao2022dkd} and exploit constant value hyperparameters $\alpha$ and $\beta$ to adjust the weight of the two components.

\section{Experiments}

\begin{table*}[!t]
\caption{Results of MKGR on DB15K, MKG-W, and MKG-Y datasets. The best and \underline{second best} results of each column are bolded and underlined, and relative improvements are listed. 
``T.'' represents the Transformer. ``Adapt.'' represents adaptive fusion and ``Select.'' represents our selective multimodal combination. 
}\label{tab: main_result}
\vspace{-10pt}
\resizebox{\linewidth}{!}{
\centering

\begin{tabular}{lcc|cccc|cccc|cccc}
\toprule
\multirow{2}{*}{Model}&\multirow{2}{*}{Backbone}&\multirow{2}{*}{Fusion} & \multicolumn{4}{c|}{DB15K} & \multicolumn{4}{c|}{MKG-W} &
\multicolumn{4}{c}{MKG-Y} \\
\cmidrule{4-7}
\cmidrule{8-11}
\cmidrule{12-15}

&&& MRR & Hits@1 & Hits@3 & Hits@10  & MRR & Hits@1 & Hits@3 & Hits@10  & MRR & Hits@1 & Hits@3 & Hits@10  \\
 \midrule
TransE   & -    & -    & 24.86 & 12.78 & 31.48 & 47.07 & 29.19 & 21.06 & 33.20 & 44.23 & 30.73 & 23.45 & 35.18 & 43.37 \\
DistMult  & -    & -    & 23.03 & 14.78 & 26.28 & 39.59 & 20.99 & 15.93 & 22.28 & 30.86 & 25.04 & 19.33 & 27.80 & 35.95 \\
ComplEx   & -    & -    & 27.48 & 18.37 & 31.57 & 45.37 & 24.93 & 19.09 & 26.69 & 36.73 & 28.71 & 22.26 & 32.12 & 40.93 \\
RotatE    & -    & -    & 29.28 & 17.87 & 36.12 & 49.66 & 33.67 & 26.80 & 36.68 & 46.73 & 34.95 & 29.10 & 38.35 & 45.30 \\
TuckER    & -    & -    & 33.86 & 25.33 & 37.91 & 50.38 & 30.39 & 24.44 & 32.91 & 41.25 & 37.05 & 34.59 & 38.43 & 41.45 \\
GC-OTE &- &- & 31.85 & 22.11 & 36.52 & 51.18 & 33.92& 26.55& 35.96 & 46.05& 32.95 & 26.77 & 36.44 & 44.08 \\ 
\midrule
IKRL & TransE    & Static & 26.82 & 14.09 & 34.93 & 49.09 & 32.36 & 26.11 & 34.75 & 44.07 & 33.22 & 30.37 & 34.28 & 38.26 \\
TBKGC & TransE    & Static & 28.40 & 15.61 & 37.03 & 49.86 & 31.48 & 25.31 & 33.98 & 43.24 & 33.99 & 30.47 & 35.27 & 40.07 \\
TransAE   & TransE    & Static & 28.09 & 21.25 & 31.17 & 41.17 & 30.00 & 21.23 & 34.91 & 44.72 & 28.10 & 25.31 & 29.10 & 33.03 \\
QEB & TransE    & Static & 28.18 & 14.82 & 36.67 & 51.55 & 32.38 & 25.47 & 35.06 & 45.32 & 34.37 & 29.49 & 36.95 & 42.32 \\
VBKGC & T.+TransE & Static & 30.61 & 19.75 & 37.18 & 49.44 & 30.61 & 24.91 & 33.01 & 40.88 & 37.04 & 33.76 & 38.75 & 42.30 \\
MACO & R-GCN & Static & 27.41 & 14.61 & 35.59 & 50.00 & 31.74 & 25.23 & 34.23 & 44.37 & 34.98 & 31.59 & 36.68 & 40.51 \\
RSME & ComplEx   & Adapt.  & 29.76 & 24.15 & 32.12 & 40.29 & 29.23 & 23.36 & 31.97 & 40.43 & 34.44 & 31.78 & 36.07 & 39.09 \\
OTKGE & ComplEx   & Adapt.  & 23.86 & 18.45 & 25.89 & 34.23 & 34.36 & 28.85 & 36.25 & 44.88 & 35.51 & 31.97 & 37.18 & 41.38 \\
MMRNS & ComplEx   & Adapt.  & 27.25 & 17.25 & 31.93 & 45.50 & 28.98 & 22.69 & 31.32 & 41.26 & 31.98 & 25.63 & 35.29 & 42.28 \\
MoSE-AI & ComplEx & Adapt. & 37.64 & 28.14 & 39.60 & 51.63 & 35.99 & 29.33 & 38.38 & 48.81 & 34.96 & 31.24 & 36.26 & 41.46 \\
MMRNS & RotatE    & Adapt.  & 29.67 & 17.89 & 36.66 & 51.01 & 34.13 & 27.37 & 37.48 & 46.82 & 35.93 & 30.53 & 39.07 & 45.47 \\
AdaMF & RotatE    & Adapt.  & 32.51 & 21.31 & 39.67 & 51.68 & 34.27 & 27.21 & 37.86 & 47.21 & 38.06 & 33.49 & 40.44 & 45.48 \\
~~~-MAT & RotatE    & Adapt.  & 35.14 & 25.30 & 41.11 & \underline{52.92} & 35.85 & 29.04 & \underline{39.01} & \underline{48.42} & \underline{38.57} & 34.34 & \underline{40.59} & \underline{45.76} \\
MMRNS & GC-OTE    & Adapt.  & 32.68 & 23.01 & 37.86 & 51.01 & 34.32 & 27.14& 36.82& 47.01& 33.38 & 27.65& 36.78& 43.44 \\
VISTA & T.  & Adapt.  & 30.42 & 22.49 & 33.56 & 45.94 & 32.91 & 26.12 & 35.38 & 45.61 & 30.45 & 24.87 & 32.39 & 41.53 \\
IMF & Tucker    & Adapt.  & 32.25 & 24.20 & 36.00 & 48.19 & 34.50 & 28.77 & 36.62 & 45.44 & 35.79 & 32.95 & 37.14 & 40.63 \\
MyGO & T.+Tucker & Adapt.  & \underline{37.72} & \underline{30.08} & \underline{41.26} & 52.21 & \underline{36.10} & \underline{29.78} & 38.54 & 47.75 & 38.44 & \underline{35.01} & 39.84 & 44.19   \\
\rowcolor{gray!10} {\textbf{DSoM}} &{\textbf{ComplEx}}  & {\textbf{Select.}}  & \textbf{42.68}  & \textbf{34.45}  & \textbf{46.62}  & \textbf{58.12} & \textbf{40.90}  & \textbf{33.56}  & \textbf{43.58}  & \textbf{55.28}  & \textbf{40.02} & \textbf{35.67} & \textbf{41.70} & \textbf{47.71} \\
\rowcolor{gray!10} \multicolumn{3}{l|}{$\Delta$ \textit{Improve.}}&   {+13.1\%} & {+14.5\%} & {+13.0\%} & {+9.8\%}&{+13.3\%} & {+12.7\%} & {+11.7\%} & {+15.8\%} & {+3.8\%} & {+1.9\%} & {+2.7\%} & {+4.3\%}\\
\midrule
KG-bert & \multirow{2}{*}{\makecell{Language\\Models}} & \multirow{2}{*}{\makecell{Adapt.\\S+D Fusion}} &  23.94 & 11.98 & 31.05 & 46.54 & 28.68 & 21.12 & 32.57 & 43.46  & - & - & - & - \\
K-ON&  &  & 38.10  & 30.13 & 42.77 & 53.59 & 36.64 & 30.05 & 38.72 & 48.26  & - & - & - & -\\
\bottomrule
\vspace{-25pt}
\end{tabular}}
\end{table*}

\subsection{Experimental Setting}
\textbf{Datasets.}
We evaluate our proposed method on 5 widely adopted MKGR datasets, including DB15K, MKG-W, MKG-Y \cite{xu2022MMRNS,zhang2024unleashing,zhang2024mygo}, and large-scale MKGs FB15K-237 and WN18 \cite{liang2023SGMPT,chen2022mkgformer}. 
We exploit the original triples, entity images, and entity descriptions provided in \cite{xu2022MMRNS,liu2019mmkgdataset,wang2021rsme,chen2022mkgformer}. 

\textbf{Evaluation. }
We evaluate the MKGR performance with widely-used evaluation settings following \cite{bordes2013TransE,wang2021rsme,xu2022MMRNS}.
We obtain the rank of the target entity throughout all the entities $\mathcal{E}$. 
Then we calculate the metrics including Mean Reciprocal Rank (MRR), Mean Rank (MR), and Hits@K (K=1,3,10). 
The MRR represents the average of the reciprocal of the target entity rank of all triples, MR represents the average of the target entity rank, while Hits@K represents the accuracy of the target entity in top K-ranked entities. 
Higher MRR and Hits@K and lower MR represent better performance. 
We conduct both head prediction and tail prediction and report their average metrics. The filter setting is applied to filter other true target entities before calculating the metrics \cite{bordes2013TransE}.

\textbf{Implementation details. }
We implement our approach with PyTorch.
We exploit ViT \cite{dosovitskiy2020ViT} (version \textit{vit-base-patch16-224}) and BERT \cite{devlin2019bert} (version \textit{bert-base-uncased}) as visual and textual feature extractors, respectively. 
The multimodal feature extraction procedures are conducted with fixed pretrained models from Transformers \cite{wolf2020huggingface}. 
The default $\gamma,\tau,\alpha,\beta$ are 2.0, 4.0, 1.0, 1.0 for all datasets. 
We exploit ComplEx \cite{trouillon2016complex} as the backbone following MoSE \cite{zhao2022mose} for a fair comparison with ensemble-based MKGR methods. 
The hidden state in the MLP of the policy network $\theta$ is 1024. 
We conduct experiments with a single NVIDIA GeForce RTX A6000 GPU with 48 GB of memory. 

\textbf{Baselines. }
We compare our method with existing MKGR approaches. 
The baselines can be roughly divided into 3 groups: \textit{(1) Unimodal KGR baselines}: TransE \cite{bordes2013TransE}, DistMult \cite{yang2015DistMult}, ComplEx \cite{trouillon2016complex}, RotatE \cite{sun2018RotatE}, Tucker \cite{balazevic2019tucker}, and GC-OTE \cite{tang2020gcote}, which are the backbone of the multimodal baselines.
\textit{(2) Multimodal KGR baselines}: 
IKRL \cite{xie2017IKRL}, TBKGC \cite{mousselly2018TBKGC}, TransAE \cite{wang2019transAE}, 
QEB \cite{wang2023QEB}, 
VBKGC \cite{zhang2022VBKGC}, 
MACO \cite{zhang2023maco}, 
RSME \cite{wang2021rsme}, OTKGE \cite{cao2022otkge}, 
MMRNS \cite{xu2022MMRNS}, 
MoSE \cite{zhao2022mose},
AdaMF and AdaMF-MAT \cite{zhang2024unleashing}, 
VISTA \cite{lee2023vista}, IMF \cite{li2023imf}, MyGO \cite{zhang2024mygo}, MKGformer \cite{chen2022mkgformer}, SGMPT \cite{liang2023SGMPT}, LAFA \cite{shang2024lafa}.
Textual KGR baselines are listed for reference since they do not model visual modality:
\textit{(3) Textual KGR baselines}: KG-bert \cite{yao2019kgbert}, SimKGC \cite{wang2022simkgc}, LMKE \cite{wang2022ijcaiLMKE}, and K-ON \cite{guo2025kon}.

\begin{table*}[!t]
\caption{Results of DSoM with teacher settings. ``ConfTeacher'' represents selecting the teacher with the largest max probability, ``BestTeacher'' and ``BestStrategy'' represent selecting the teacher or strategy with the lowest cross-entropy.
}\label{tab: ablation_kd}
\vspace{-10pt}
\resizebox{\linewidth}{!}{
\centering
\small

\begin{tabular}{l|cccc|cccc|cccc}
\toprule
\multirow{2}{*}{Model}&\multicolumn{4}{c|}{DB15K} & \multicolumn{4}{c|}{MKG-W} &
\multicolumn{4}{c}{MKG-Y} \\
\cmidrule{2-5}
\cmidrule{6-9}
\cmidrule{10-13}
& MRR & Hits@1 & Hits@3 & Hits@10  & MRR & Hits@1 & Hits@3 & Hits@10  & MRR & Hits@1 & Hits@3 & Hits@10  \\
 \midrule



{\textbf{DSoM}}& \textbf{42.68}  & \textbf{34.45}  & \textbf{46.62}  & \textbf{58.12} & \textbf{40.90}  & \textbf{33.56}  & \textbf{43.58}  & \textbf{55.28}  & \textbf{40.02} & \textbf{35.67} & \textbf{41.70} & \textbf{47.71}\\

\midrule

\quad w/ ConfTeacher    & 40.87 & 32.82 & 44.75 & 56.16 & 37.49 & 31.50 & 39.64 & 48.97 & 37.76 & 33.93 & 39.82 & 44.24 \\
\quad w/ BestTeacher  & 41.43 & 33.46 & 45.40 & 56.56 & 38.27 & 32.41 & 40.48 & 49.61 & 38.51 & 34.89 & 40.39 & 44.71 \\
\quad w/ BestStrategy  & 41.65 & 33.60 & 45.81 & 56.94 & 39.14 & 32.88 & 41.55 & 51.10 & 39.34 & 35.58 & 41.01 & 45.74 \\
\quad w/ MulDE    & 39.94 & 31.74 & 43.52 & 56.03 & 38.68 & 32.39 & 41.11 & 50.36 & 39.62 & 35.60 & 41.19 & 46.68 \\

\bottomrule
\end{tabular}}
\vspace{-10pt}
\end{table*}
\begin{table}[!t]
\caption{Results of MKGR on FB15K-237 and WN18 datasets. The baseline results are from \cite{liang2023SGMPT,sun2018RotatE,shang2024lafa}.
}\label{tab: fbwn}
\vspace{-10pt}
\resizebox{\linewidth}{!}{
\centering
\small

\begin{tabular}{l|cccc|cccc}
\toprule
\multirow{2}{*}{Model}&\multicolumn{4}{c}{FB15K-237}&\multicolumn{4}{|c}{WN18} \\
\cmidrule{2-5}
\cmidrule{5-9}
& MR & H@1 & H@3 & H@10 & MR & H@1 & H@3 & H@10   \\
 \midrule
IKRL & 298 & 19.4 & 28.4 & 45.8 & 596 & 12.7 & 79.6 & 92.8 \\
TransAE & 431 & 19.9 & 31.7 & 46.3 & 352 & 32.3 & 83.5 & 93.4 \\
RSME & 417 & 24.2 & 34.4 & 46.7 & 223 & 94.3 & 95.1 & 95.7 \\
MoSE-AI & \underline{135} & 25.5 & 37.6 & 51.8 & \underline{23}  & 92.9 & 94.6 & 96.2 \\
MyGO & & 19.0 & 28.9 & 44.7 & - & 70.6 & 93.7 & 94.1 \\
AdaMF-MAT   & -   & 23.1 & 35.0 & 49.1 & - & 73.6 & 94.3 & 95.8 \\
MKGformer   & 252 & 24.3 & 36.0 & 49.9 & {25}   & 93.5 & 95.8 & 97.0 \\
SGMPT   & 238 & 25.2 & 37.0 & 51.0 & 29  & 94.3 & \textbf{96.6} & \textbf{97.8} \\
LAFA & {136}    & \underline{26.9}    & \underline{39.8}    & \underline{55.1}    & {25}   & \underline{94.5}    & \underline{96.2} & 97.3 \\
\rowcolor{gray!10} \textbf{DSoM} & \textbf{116} & \textbf{27.5} & \textbf{41.1} & \textbf{56.4} & \textbf{7} & \textbf{95.0} & \underline{96.2}    & \underline{97.4}   \\
\rowcolor{gray!10} $\Delta$ \textit{Improve.}& 13.8 \% & 2.2\% & 3.3\% & 2.4\% & 69.6\% & 0.5\% & - & -\\
\midrule
KG-bert & 153  & - & - & 42.0  & 58 & 11.7  & 68.9  & 92.6 \\
SimKGC  & -   & 24.6 & 36.2 &51.0    & - & 90.2 & 94   & 96.1 \\
LMKE & -   & 21.5 & 33.0 & 48.1 & - & - &-  &-  \\
\bottomrule
\end{tabular}}
\vspace{-20pt}
\end{table}

\subsection{Main Results}
Table \ref{tab: main_result} shows the MKGR performance of DSoM and baselines on DB15K, MKG-W, and MKG-Y datasets, and Table \ref{tab: fbwn} shows performance on FB15K-237 and WN18. 
DSoM outperforms the baselines in Table \ref{tab: main_result} by over $10\%$ on 7 metrics, and outperforms the baselines in Table \ref{tab: fbwn} on 6 metrics. 
It shows the significant improvements of DSoM over baselines. 

\textbf{Comparing Unimodal KGR baselines}, DSoM significantly outperforms them, especially its backbone ComplEx \cite{trouillon2016complex}, which demonstrates the effectiveness of DSoM in exploiting the reinforced combined multimodal soft labels to enhance the unimodal student model. 
\textbf{Compared with Multimodal KGR baselines}, DSoM also outperforms all of them, even some of which exploit stronger backbones such as RotatE, Tucker, GC-OTE, and Transformers. 
It also demonstrates the effectiveness of our proposed methods. 
DSoM also outperforms the ensemble-based methods MoSE and IMF, demonstrating the effectiveness of our distillation-based optimization and reinforced multimodal selection strategies. 
\textbf{Compared with Textual KGR baselines}, even without their strong pre-trained language models, DSoM outperforms them with a simple embedding-based backbone, demonstrating our effectiveness. 
Moreover, they did not exploit visual modality, which could also be the reason.

\textbf{Comparing static- and adaptive-fusion baselines}, DSoM outperforms both of them significantly. 
In previous studies, most adaptive-based baselines outperform the static-based ones, since they assign dynamic weights to multimodal information, thus better facilitating the multimodal complements. 
In contrast, DSoM thoroughly prevents the unhelpful modalities for each triple in model optimization through reinforced multimodal combination, which shows promising performance.
Moreover, a naive mean aggregation of selected teachers in DSoM can outperform baselines.
It further emphasizes the potential of excluding the unhelpful modalities rather than adaptive fusion. 
It also leaves the dynamic aggregation of the selected teachers for future work. 

\textbf{Performance on large-scale MKGs FB15K-237 and WN18.}
DSoM outperforms most baselines on large-scale MKGs, especially on MR metric. 
Since MR represents the stability of the framework, it shows the promising scalability of DSoM. 
DSoM reaches comparable Hit@3 and Hit@10 performance on WN18 with SGMPT. 
The reason could be that WN18 has unremoved reverse triples \cite{dettmers2018conve-wn18rr}, thus it is easy to reach over $90\%$ performance.
It also made it hard to reach higher performance than transformer-based architectures \cite{liang2023SGMPT} with our embedding-based framework.

\subsection{Ablation Study}
Table \ref{tab: ablation} shows the ablation study of DSoM. 
The Teacher Avg. represents the average performance of three modal teachers, which is the baseline of our framework. 

\textbf{Effectiveness of RC and NDKD. }
We first add RC and NDKD separately to Teacher Avg. to demonstrate their effectiveness. 
We conduct the multimodal reinforcement combination learning on the validation set and then transfer to the test set, comparing the meta-learner in MoSE \cite{zhao2022mose}. 
The adoption of our reinforced combination strategy outperforms both the Teacher Avg. and meta-learner \cite{zhao2022mose}, demonstrating the effectiveness of the reinforced multimodal selection. 
Then, we directly distill the Teacher Avg. with our proposed NDKD to a unimodal student model without reinforced combination, which outperforms the Teacher Avg. and demonstrates the effectiveness of exploiting dark knowledge in soft labels. 

\begin{table}[!t]
\caption{Ablation study on DB15K. ``RC'' and ``NDKD'' denote Reinforced Combination and Neighbor-Decoupled KD. ``ML'' denotes meta-learner in MoSE \cite{zhao2022mose}. 
$\mathcal{L}_{NEKD}$, $\mathcal{L}_{NNKD}$, and DKD are in Equation \eqref{eq:dkd}, Vanilla KD in Equation \eqref{eq: KD}.
}\label{tab: ablation}
\vspace{-10pt}
\resizebox{0.9\linewidth}{!}{
\centering
\small

\begin{tabular}{l|cccc}
\toprule

Model & MRR & Hits@1 & Hits@3 & Hits@10   \\
 \midrule

Teacher Avg.    & 37.64 & 28.14 & 39.60 & 51.63 \\
\midrule
\quad w/ ML on val. \cite{zhao2022mose}&40.62 & 32.34 & 44.53 & 55.29 \\
\quad w/ RC on val. & 41.08 & 33.21 & 45.01 & 56.08   \\
\quad w/ NDKD   & 41.91 & 33.69 & 45.83 & 57.56\\
\midrule
\quad w/ RC \& $\mathcal{L}_{NEKD}$ &41.00 & 33.07 & 44.93 & 56.00\\ 
\quad w/ RC \& $\mathcal{L}_{NNKD}$ &42.33 & 33.93 & 46.39 & 58.08   \\ 
\quad w/ RC \& Vanilla KD & 42.03 & 33.62 & 46.07 & 57.96 \\
\quad w/ RC \& DKD & 42.31 & 34.00	& 46.16	& 58.07 \\
\midrule
\textbf{DSoM}   & \textbf{42.68}  & \textbf{34.45}  & \textbf{46.62}  & \textbf{58.12}  \\
\bottomrule
\end{tabular}}
\vspace{-10pt}
\end{table}

\textbf{Ablation of NDKD. }
We added the RC and conducted the ablation study of NDKD. 
Adding $\mathcal{L}_{NEKD}$ and $\mathcal{L}_{NNKD}$ separately both benefits the student model. 
Moreover, the non-neighbor distillation also outperforms the vanilla KD, showing that even only considering the non-target entity correlations in KD would benefit the model. 
Since the non-neighbor entities are the majority of dark knowledge, the performance of involving $\mathcal{L}_{NNKD}$ outperforms that of $\mathcal{L}_{NEKD}$.
We also compare our method with the decoupled KD \cite{zhao2022dkd} which only decouples the target entities without considering other neighbors. 
The NDKD of DSoM outperforms DKD, demonstrating the necessity of considering all true targets. 

\subsection{Effect of Distillation Methods}
Table \ref{tab: ablation_kd} shows the performance of DSoM by replacing our distillation mechanisms with other strategies. 

\textbf{Comparison with teacher selection strategies. }
We first construct the multimodal teacher selection mechanisms, including ConfTeacher, BestTeacher, and BestStrategy, to replace our reinforced mechanism. 
The experiment results show that our reinforced teacher combination mechanism outperforms the heuristic teacher selection strategies, even the greedily computed best strategy.
This is possible because our proposed reinforcement learning could find the optimal set of multimodal teachers that are beneficial for the student. This strategy may not be the best strategy itself, but it is the best for the student. 
It is reasonable since, as the saying ``\textit{{Teaching students in accordance with their aptitude}}'', it is more important to consider what is in the best interest of the student model, which our proposed reinforced mechanism could autonomously achieve. 

\textbf{Comparison with multi-teacher KD. }
We adopt the multi-teacher knowledge distillation framework for KGR named MulDE \cite{wang2021MulDE} to DSoM. 
MulDE proposes a Senior model to adaptively integrate the multiple teachers based on relation-specific scaling and contrast attention to teach to the smaller Junior model. 
We reimplemented MulDE to distill the multimodal teachers to the student model. 
The results show that DSoM outperforms MulDE. 
Though MulDE outperforms MoSE-AI, demonstrating the effectiveness in their relation-specific modality adaptive fusion, DSoM still outperforms MulDE for excluding the negative impacts of unhelpful modalities.
Moreover, MulDE also outperforms existing MKGR methods, which further validates our hypothesis in emphasizing dark knowledge in non-target entities through knowledge distillation for MKGR. 


\begin{figure}[t]
\centering
\subfigure[Strategy ratio in 3 datasets.]
{
\label{fig:strategy_dataset}
\includegraphics[width=0.45\textwidth]{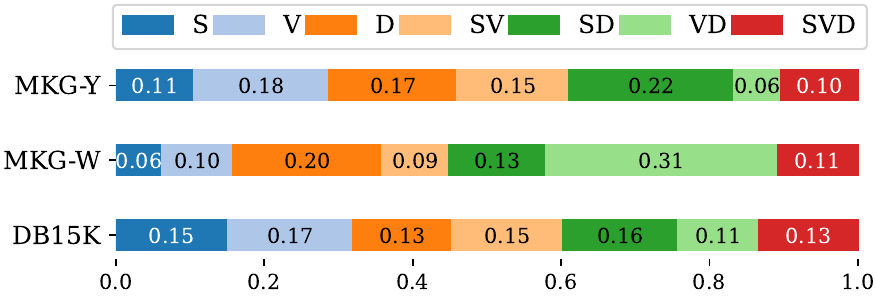}}
\subfigure[Strategy ratio in triples consists of specific entity/relation.]
{
\vspace{-40pt}
\label{fig:strategy_ent_rel}
\includegraphics[width=0.45\textwidth]{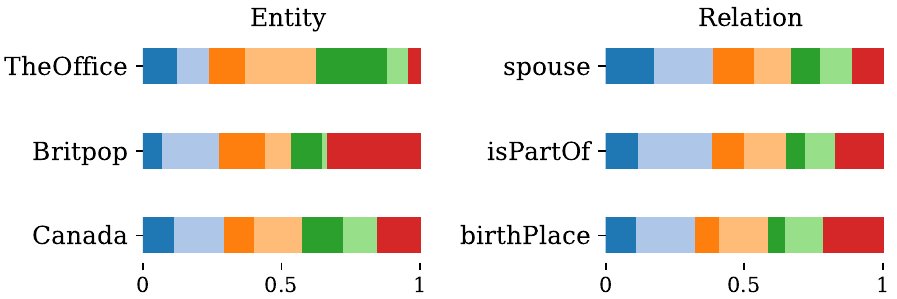}}
\vspace{-15pt}
\caption{The ratio of selected strategies of DSoM. The S, V, and D represent structural, visual, and textual modalities. }\label{fig:strategy}
\vspace{-18pt}
\end{figure}

\subsection{Reinforced Selected Strategy Statistics}
Figure \ref{fig:strategy} shows the ratio of selected multimodal combination strategies of three datasets and the triples consist of specific entities or relations. 
As shown in Figure \ref{fig:strategy}, the triples in different KGs, consisting of different entities and relations, could dynamically select the optimal set of multimodal teachers. 
Moreover, since six out of seven strategies include visual or textual modalities, it demonstrates the benefits of including more modalities for KGR, which provides more semantic information and improves KGR performance. 
Moreover, based on the strategy ratio analysis, the MKG management could purposefully improve the modality quality or knowledge coverage of certain modalities, to further improve the quality of MKG.

\subsection{Efficiency Analysis}
\begin{figure}[!t]
    \centering
    \includegraphics[width=0.48\textwidth]{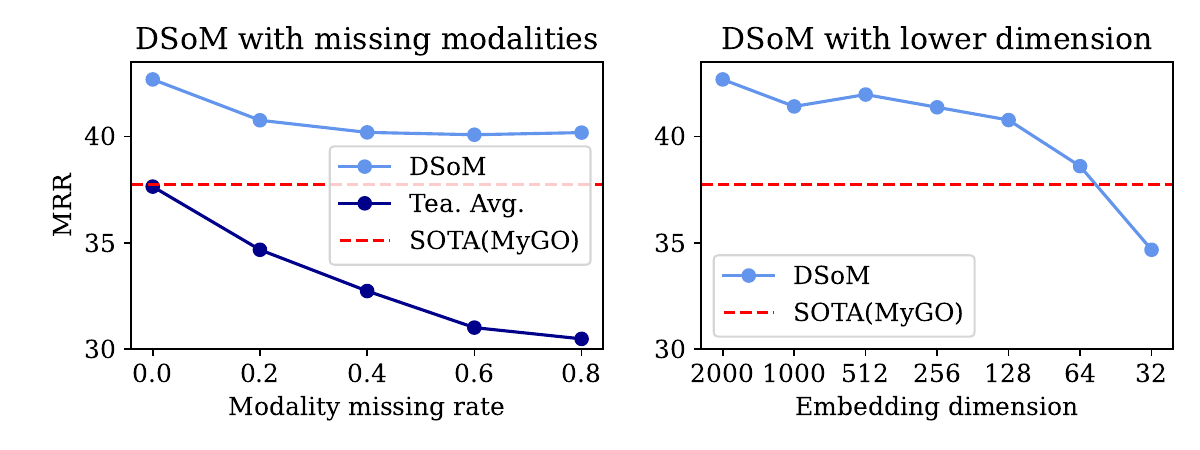}
    \vspace{-15pt}
    \caption{Efficiency analysis on DB15K of DSoM with missing modalities and lower embedding dimension.}
    \label{fig:efficiency}
    \vspace{-25pt}
\end{figure}

Since the knowledge distillation are widely used in model compression \cite{hinton2015KD}, as well as KGR model compression \cite{wang2021MulDE}, we also study the efficiency of DSoM. 
Figure \ref{fig:efficiency} shows the efficiency of DSoM with missing modalities and with lower embedding dimensions. 

\textbf{Effectiveness with missing modalities. }
By increasing the modality missing rate (we set the same rate for visual and textual modalities \cite{zhang2024unleashing}), the Teacher Avg. shows rapid down-gradation, while DSoM shows smooth decreasing performance and always outperforms the SOTA model MyGO with full modalities. 
The reason is that with the reinforced combined multimodal knowledge distillation, the student model only learns from helpful teachers.
With more missing modalities, as the saying goes ``\textit{{select the essence and discard the gross}}'', the student would only learn from the good essence and eliminate the noisy and negative teachers. 

\textbf{Effectiveness with less trainable parameters. }
By decreasing the embedding dimension, the DSoM shows a limited decrease until the dimension is under 128. 
Moreover, the trainable parameters of MyGO are $22.0M$, which is between the $26.8M$ of DSoM with $dim=1000$ and $13.72M$ of DSoM with $dim=512$. 
DSoM consistently outperforms SOTA with larger or smaller embedding dimensions. 
Moreover, DSoM with $dim=64$ has only $1.72M$ trainable parameters and shows $38.60$ MRR, i.e., \textbf{DSoM with $7.8\%$ of MyGO parameters outperforms MyGO by $2.3\%$ in MRR}.
It shows both the effectiveness and efficiency of DSoM.
Moreover, {the training time of DSoM is merely 10 seconds per epoch} (DB15K), while those of MyGO, MMRNS, and OTKGE are 11s, 23s, and 64s, \cite{zhang2024mygo}, respectively. 
It shows that our proposed neighbor-decoupled KD fully unleashes the potential by exploiting the dark knowledge in non-target entities to provide rich supervision signals.

\section{Conclusion}

In this paper, we explore the Dark Side of Modalities for MKGR in two folds: 
To capture the dark knowledge in non-target entities, we propose a neighbor-decoupled knowledge distillation method to train a student model with multimodal soft labels, capturing the entity label correlations in multimodal teacher logits. 
To eliminate the negative impacts of unhelpful modalities, we propose a reinforced multimodal combination mechanism to dynamically select a beneficial set of multimodal teachers for the student model. 
Experiments demonstrate the effectiveness and efficiency of our proposed framework on 5 datasets, and validate the necessity of exploiting dark knowledge and excluding unhelpful modalities. 
In the future, exploring (1) multi-agent reinforcement learning with a more comprehensive multimodal environment, and (2) adaptive multimodal teacher aggregation considering teacher-student correlations would be promising to improve the effectiveness and efficiency of MKGR.



\section*{Acknowledgements}
This research is supported by the National Natural Science Foundation of China (No. 62272250) and the Natural Science Foundation of Tianjin, China (No. 22JCJQJC00150).

\bibliographystyle{ACM-Reference-Format}
\bibliography{mybib}


\newpage

\appendix

\newpage
\begin{algorithm}[!t]
    \SetKwInOut{Input}{Input}
    \SetKwInOut{Output}{Output}
    \SetKwInOut{Require}{Require}
    
    \Input{MKG $\mathcal{G}=\{\mathcal{E},\mathcal{R},\mathcal{T},\mathcal{V},\mathcal{D}\}$
    }
    \Output{Student KGE model $\mathbf{E}^{stu},\mathbf{R}^{stu}$}
{
    // Multi-modal Teachers Pre-training\\
    \For{each epoch}{\For{$(h,r,t) \in \mathcal{T}_{train}$}{
    Optimize multimodal teacher KGE models $\mathbf{E}^m, \mathbf{R}^m,m\in\{\mathcal{S,V,D}\}$ as Equation (3).
    }}
    // Student Training \\
    \For{each epoch}{\For{$(h,r,t) \in \mathcal{T}_{train}$}{
    Construct state as Equation (5).\\
    Sample multimodal teacher combination strategy as Equation (6).\\
    Aggregate teacher logits as Equation (7). \\
    Sample neighbors $N(h,r)$ in training graph $\mathcal{T}_{train}$.\\
    Temperature-scaling and neighbor-decoupling of teacher logits and student logits. \\
    Optimize student model $\mathbf{E}^{stu},\mathbf{R}^{stu}$ and policy network $\theta$ as Equation (4). 
    }}
    // Student Evaluation \\
    \For{$(h,r,t) \in \mathcal{T}_{test}$}{
    Calculate score distributions with student model and score function as $\mathbf{F}(h,r,t)=[f(h,r,e;\mathbf{E}^{stu},\mathbf{R}^{stu})]\in \mathbb{R}^{|\mathcal{E}|}, e \in \mathcal{E}$.
    }
}
    \caption{Pre-training, Training, and Testing of DSoM}
    \label{alg: pipeline}
\end{algorithm}

\section{Pipeling of DSoM} \label{app:pipeline}
The pipeline of DSoM, including multimodal teacher pre-training, student training, and student evaluation, is shown in Algorithm \ref{alg: pipeline}. We will open-source our code on GitHub after the paper is accepted.

\section{Time Complexity Analysis}

\textbf{(1) Linear Time Complexity:} The reinforcement selection and pre-training steps all require linear time complexity with the training set size, since
- The multi-modal teachers are simultaneously multi-task trained, where the backbone ComplEx remains linear time complexity.
- State, action, and reward are all linear operations completed in a single step.

\textbf{(2) Time and Parameter Efficiency}: Our pre-training is the same as training of MoSE, while our reinforced distillation can be seen as ensemble of MoSE; thus, DSoM does not take extra steps than the ensemble-based baseline. 

On DB15K, pre-training requires 17 s/epoch with 64.2M trainable parameters, while reinforcement selection and distillation require 10 s/epoch with 53.6M trainable parameters. The efficiency is reasonable considering the performance increase. 

\textbf{(3) Comparison to SOTA:} The efficiency analyses are detailed in Section 5.6, where DSoM outperforms SOTA with fewer available modalities (even 80\% missing) and fewer trainable parameters (DSoM with 7.8\% parameters of SOTA outperforms by 2.3\% in MRR).

\begin{table}[!b]
    \caption{Dataset statistics of MKGR.}

    \begin{center}
    
    \resizebox{0.4\textwidth}{!}{
    \small
\begin{tabular}{l|rrrrr}
\toprule
Dataset   & $|\mathcal{E}|$ & $|\mathcal{R}|$ &  Train & Valid & Test \\
\midrule
DB15K & 12,842    & 279 & 79,222 & 9,902 & 9,904  \\
MKG-W      & 15,000     & 169 & 34,196 & 4,276   & 4,274  \\
MKG-Y       & 15,000      & 28 & 21,310 & 2,665   & 2,663\\
FB15K-237&14,541&237&272,115& 17,535& 20466\\
WN18&40,943&18& 141,442& 5,000& 5,000\\
WN9 & 6,555&9&11,741 & 1,337 & 1,319 \\
\bottomrule
    
    \end{tabular}}
    \end{center}
    \label{tab: datasets} 
    \end{table}

\begin{table*}[!t]
\caption{Results of DSoM with modality ablation. The ``S/V/D'' represents structural, visual, and textual teacher, respectively.
}\label{tab: ablation_mod}
\resizebox{\linewidth}{!}{
\centering
\small

\begin{tabular}{l|cccc|cccc|cccc}
\toprule
\multirow{2}{*}{Model}&\multicolumn{4}{c|}{DB15K} & \multicolumn{4}{c|}{MKG-W} &
\multicolumn{4}{c}{MKG-Y} \\
\cmidrule{2-5}
\cmidrule{6-9}
\cmidrule{10-13}
& MRR & Hits@1 & Hits@3 & Hits@10  & MRR & Hits@1 & Hits@3 & Hits@10  & MRR & Hits@1 & Hits@3 & Hits@10  \\
 \midrule

{\textbf{DSoM}}& \textbf{42.68}  & \textbf{34.45}  & \textbf{46.62}  & \textbf{58.12} & \textbf{40.90}  & \textbf{33.56}  & \textbf{43.58}  & \textbf{55.28}  & \textbf{40.02} & \textbf{35.67} & \textbf{41.70} & \textbf{47.71}\\
\midrule
\quad w/ S teacher & 41.34 & 33.46 & 45.29 & 56.39 & 36.81 & 31.02 & 38.79 & 47.99 & 38.33 & 34.92 & 40.14 & 44.24 \\
\quad w/ V teacher & 37.36 & 30.19 & 40.92 & 51.03 & 31.18 & 25.92 & 33.53 & 40.84 & 33.92 & 30.79 & 35.09 & 39.05 \\
\quad w/ D teacher & 41.30 & 32.87 & 45.25 & 57.48 & 39.23 & 31.50 & 41.94 & 54.05 & 37.29 & 32.73 & 39.24 & 45.16 \\
\bottomrule
\end{tabular}}
\end{table*}

\section{Dataset Statistics}
We conduct experiments with 6 MKGR datasets, including DB15K, MKG-W, MKG-Y from \cite{xu2022MMRNS}, FB15K-237, WN18 from \cite{wang2021rsme}, and WN9 from \cite{xie2017IKRL}. 
The DB15K dataset \cite{liu2019mmkgdataset} is from DBPedia KG \cite{lehmann2015dbpedia}, the MKG-W and MKG-Y datasets are constructed by \cite{xu2022MMRNS} from Wikipedia \cite{vrandevcic2014wikidata} and YAGO \cite{suchanek2007yago}, FB15K-237 dataset \cite{toutanova2015FB15k237} is from Freebase KG \cite{bollacker2008freebase}, and WN18, WN9 dataset is from WordNet KG \cite{Miller1995wordnet}, respectively. 
The dataset statistics are listed in Table \ref{tab: datasets}.

\section{Experiments compared with Structural MKGR methods}

We compare DSoM with TMR \cite{zheng2025TKDE_MKGR}, DySarl \cite{liu2024dysarl}, and MMKGR \cite{zheng2023icde_mkgr} on their common dataset, WN9, since they do not use our datasets. The results demonstrate the effectiveness of DSoM compared with them. MRR, Hits@1 of DSoM outperforms baselines by 3.9\% and 8.1\%, respectively. DSoM did not outperform Hits@10 of DySarl, possibly because DSoM did not exploit strong structural learning networks. In this paper, we mainly focus on multimodal learning of MKGR; thus, we exploit a basic KGE model, ComplEx, as our backbone. We will add the discussions in the final version and explore deep structural learning for DSoM in future work. 

\begin{table}[!t]
\caption{Experiments on WN9 compared to deep structure based methods. }\label{tab: wn9}
\resizebox{0.7\linewidth}{!}{
\centering
\small

\begin{tabular}{l|ccc}
\toprule

Model & MRR & Hits@1  & Hits@10   \\
 \midrule

MKGR \cite{zheng2023icde_mkgr}   & 80.20  & 73.60  & 92.80   \\
TMR \cite{zheng2025TKDE_MKGR}   & 86.30  & 79.70  & 93.70   \\
DySarl \cite{liu2024dysarl} & 88.65 & 83.70  & \textbf{96.40}   \\
DSoM   & \textbf{92.13} & \textbf{90.52} & 95.56  \\
$\Delta$  & 3.9\% & 8.1\% & -0.9\% \\
\bottomrule
\end{tabular}}
\end{table}

\section{Modality Ablation}

We conduct a modality ablation study with the single-modal teacher for DSoM training. As shown in Table \ref{tab: ablation_mod}, DSoM outperforms all single-modal teacher settings, showing the effectiveness of DSoM in combining the multimodal teachers. 
Moreover, Table \ref{tab: ablation_mod} also shows that structural modality and textual modality are the most helpful teachers, while the visual modality is less effective. 
By dynamically combining effective teachers and eliminating unhelpful teachers in a sample-wise granularity, DSoM achieves superior performance. 

\section{Hyperparameter Analysis} \label{sec: params}
\begin{figure}[t]
    \centering
    \includegraphics[width=0.48\textwidth]{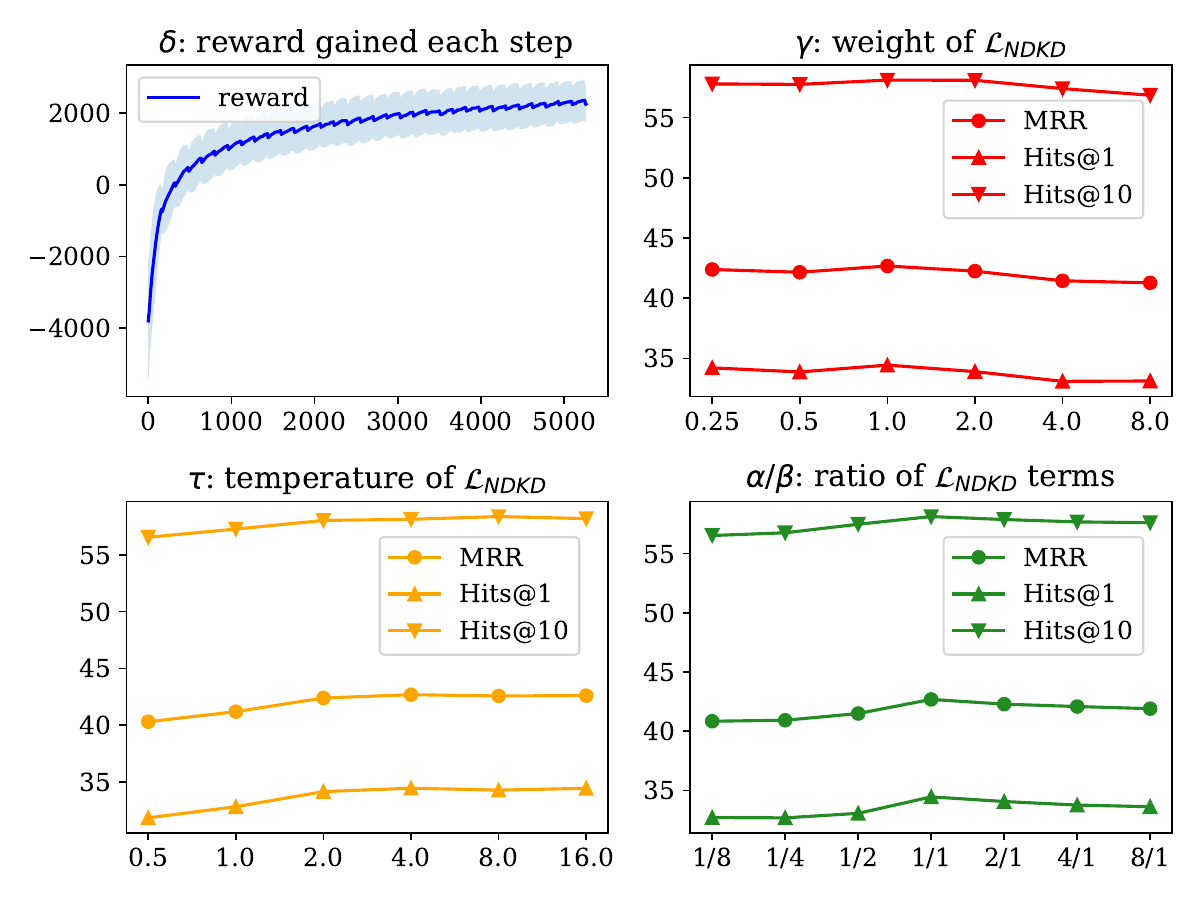}
    \vspace{-15pt}
    \caption{Reward curve $\delta$ during training, and hyper-parameter analysis of $\gamma$, $\tau$ and $\alpha/\beta$ on DB15K dataset.}
    \label{fig:params}
\end{figure}

Figure \ref{fig:params} shows the reward during training and the performance of DSoM with different hyperparameters. 
The reward of DSoM consistently increases during training steps, illustrating the agent's progressive process of finding the optimal set of teachers for the student. 
We conduct hyperparameter analysis for the hyperparameters $\gamma,\tau$, and $\alpha/\beta$. 
$\gamma$ controls the ratio of the distillation objective.
By increasing $\gamma$, the student model learns more from multimodal soft labels and less from hard labels. 
DSoM performance first increases, then decreases, showing the importance of balancing hard labels and soft labels. 
With larger $\tau$, the teacher logits get smoother and the student model gets more entropy in non-target entity labels and learns more from their correlations.
Thus, DSoM performance increases with larger $\tau$ and then converges.
$\alpha/\beta$ controls the proportion between neighbor entities and non-neighbor entities in $\mathcal{L}_{NDKD}$. 
It shows that balancing two correlations with $\alpha/\beta$ in distillation could bring better performance.

\end{document}